\documentstyle[aps,multicol,amsfonts,amssymb,amsmath,graphicx]{revtex}

\makeatletter
\def\thepage{\@arabic\c@page}
\def\@pnumwidth{2em}
\makeatother

\begin{document}

\draft
\title{\huge Neutrino decay and long base-line oscillation experiments}
\author{Vo Van Thuan$^*$, Nguyen Tuan Anh and Le Huu Thang}
\address{Institute for Nuclear Science and Technique - VINATOM,\\
Hoang Quoc Viet Street, Nghiado, Cau-Giay, Hanoi, Vietnam. \\
{\footnotesize $^{*)}$Email: vkhkthn@netnam.org.vn} }
\date{Feb. 9, 2000}
\maketitle

\begin{abstract}
Considering neutrinos as time-like leptons one may estimate the
three-body decay probability of muon neutrinos in long base-line
accelerator experiments. In the extreme assumption of time-space
symmetry the absolute value of the transcendent mass of a  muon
neutrino is equal to the rest mass of its bradyon  partner which
is, however, strongly suppressed in measurements using the weak
interaction. This decay,  neglecting small oscillations or other
effects, leads to a strong dependence of the effect on the
base-line distance. As a result, few hundred kilometre long
baseline experiments might hardly  see muon-like events. Total
rates of electron-like events from three-body decay are calculated
for K2K, MINOS and ICARUS. Shorter base-line experiments able to
see clearly  the effect of muon neutrino decay are very promissing
component of long base-line projects.\\ \vspace{0.1cm}

{\footnotesize To be submitted to the Workshop on "Neutrino
oscillations and their origin", Fuji-Yoshida, Japan, 11-13
February 2000.}
\end{abstract}


\makeatletter \global\@specialpagefalse
\def\@oddhead{V.V.Thuan at al.\hfill Neutrino decay...}
\let\@evenhead\@oddhead
\def\@oddfoot{\reset@font\rm\hfill \thepage\hfill
\ifnum\c@page=1
  \llap{\protect\copyright{} 2000}%
\fi } \let\@evenfoot\@oddfoot \makeatother


\begin{multicols}{2}
\section{Introduction}

The solar neutrino problem and the atmospheric anomaly belong to
the category of  long base-line phenomena using natural neutrino
sources. Along with the short base-line oscillation experiments
the long base-line projects with man-made neutrino sources are
very promising. Among them the recent Chooz project using reactor
neutrinos was completed with a negative effect \cite{apol}. Most
others are planned for installation at high energy accelerators.
The K2K project \cite{jung} has taken the first run recently and
has shown a preliminary first cc-event \cite{suzu}. MINOS at
Fermilab-Soudan \cite{mich} and  ICARUS at CERN-Gran Sasso
\cite{rubb} are in a preparatory  stage.

The traditional interpretation of the origin of  neutrino
anomalies is the oscillation mechanism. Another approach is the
possibility of decay of conventional massive neutrinos, which may
decay in some mixing components into a singlet majoron and another
neutrino, as summarised in Ref.~\cite{barg}. This decay theory is
restricted by very low upper limits on neutrino masses.

As an alternative, we suggested in Ref.~\cite{bloi} a three-body
decay mechanism to explain the existing oscillation hints, which
considers neutrinos as time-like leptons and where the heavier
neutrinos may decay according to  the dual principle in the
Super-luminous Lorentz Transformation (SLT).

\section{Tachyon and neutrino decay}

The idea is based on the symmetry between space-like and time-like
bradyons, following E. Recami et al. \cite{reca}, according to
which we may suggest that in the Super-luminous Lorentz
transformation the space and the time dimensions should replace
each other. As a result, tachyons may travel in a 3-dimension time
while moving in an unique space direction. A very severe problem
is that no  tachyon was ever observed in an experiment. On the
other hand, neutrinos are an exception with the following
appearances of tachyon properties:

- There is a very high symmetry between neutrinos and space-like
leptons which is enhanced in the electro-weak unification;

- Each neutrino has its unique space direction, left or right,
described by a definite helicity;

- Neutrinos never stop in a space position, similarly as a
space-like particle can never stop in a moment of time-evolution.

A big challenge is the fact that all neutrinos seem to have very
small mass, which disturbs the lepton-neutrino symmetry. To solve
this problem we assumed in Ref.~\cite{bloi} that neutrinos are
realistic tachyons, however due to weak interaction their
transcendent masses m being complex have to be strongly
suppressed. We suggested that the real part of mass is roughly
equal to $\Gamma/2$, where $\Gamma = 2\rho^2 m_0 = 192^{-1}
\pi^{-3} G_F^2 m_0^5$ is the decay width of the unstable lepton
and $m_0$ is its rest mass. The imaginary part of tachyon mass is
suppressed by the factor $\rho$, i.e. to the first order by the
Fermi constant $G_F$. This means that the imaginary part may be
measured in parity non-conserving experiments by interference of
the weak interaction and electro-magnetic or nuclear force.
Generally, we may  vary  the rest mass $m_0$ as a free parameter
to fit the experiments.  In this work,  however, we prefer to test
the first order approximation by  making the extreme assumption of
time-space symmetry, i.e. that the  absolute value of the
transcendent mass of a neutrino is equal to the rest mass of its
bradyon partner. For unification of the formalism we extend a
similar formula for $\Gamma$ of unstable leptons to the electron,
which does not harm the conclusion that electron neutrinos as
time-like electrons should be stable. As a result, we have got a
formalism almost without free parameters in our calculation. In
Table \ref{tab:mass} we show the observable transcendent masses of
space-like neutrino-tachyons or time-like leptons:
\end{multicols}
\rule{8.4cm}{.1mm}\rule{-.1mm}{.1mm}\rule{.1mm}{2mm}
\begin{table}[h]
\caption{Observable transcendent masses of neutrino-tachyons
\label{tab:mass}}
\begin{center}
\begin{tabular}{cccc}
lepton-bradyon & electron & $\mu$-meson & $\tau$-lepton
\\ \hline
$m_0$, MeV & $0.51$ & $105.6$ & $1777$ \\ $\tau$, sec & $> 1.36\
10^{31}$ & $2.2\ 10^{-6}$ & $2.9\ 10^{-13}$
\\
$\Gamma$, eV & $< 5.\ 10^{-22}$ $^{(\diamond )}$ & $2.5\ 10^{-10}$
& $1.9\ 10^{-3}$ \\ Re$(m^{\infty })$, eV & $\leq 2.5\ 10^{-22}$ &
$1.25\ 10^{-10}$ & $9.5\ 10^{-4}$ \\ $-$ Im$(m^{\infty })^{2}$,
eV$^{2}$ & $1.2\ 10^{-16}$ & $1.34\ 10^{-2}$ & $1.7\ 10^{6}$ \\
$\rho^{2}$ & $5.\ 10^{-28}$ & $1.2\ 10^{-18}$ & $5.3\ 10^{-13}$
\end{tabular}
\end{center}
$^{(\diamond )}$ {\footnotesize a similar formula for $\Gamma$ of
unstable leptons is extended to electron.}
\end{table}
\hspace{9cm}\rule{.1mm}{-4mm}\rule{.1mm}{.1mm}\rule{8cm}{.1mm}
\begin{multicols}{2}
In principle, the real part of neutrino masses may cause
oscillation effects, however, they are too small to be seen in
current experiments. The negligible transcendent masses lead to an
experimental fact that neutrinos are always identified as luxons,
i.e. particles moving with a speed of light. Now in accordance
with the dual principle in SLT, we may consider the muon neutrino
as a time-like muon, which is able to decay similarly to the decay
of a conventional space-like muon. The decay scheme is:
\begin{equation}
\nu _{\mu }\rightarrow \nu _{e}+\mu +e; \label{munud}
\end{equation}

At variance with the conventional muon decay, due to the energy
conservation, the process (\ref{munud}) takes off at a threshold
$E_\nu = 106.1$ MeV, similarly to pair production of a gamma ray.
It means that muon neutrinos do not decay at rest (DAR) and we are
able to observe only the neutrino decay in flight (DIF). In
general, we can not warrant the conservation of lepton charge in
the decay (\ref{munud}) as neutrinos and leptons may be Majorana
ones.

In the three-body decay of a muon neutrino, we may suggest that
the muon as well as the positron leave each other  from a very
short distance ($\leq 10^{-16}$ cm), where the weak interaction
dominates the Coulomb attractive force and that they may exist in
electric charge  (or lepton charge) mixing states as Majorana
particles:
\[
|\phi^0_{\mu }(\mbox{Left})\rangle =\cos \theta_{\mu} |\mu
^{+}\rangle +\sin \theta_{\mu} |\mu ^{-}\rangle
\]
\[
|\phi^0_{e}(\mbox{Right})\rangle =-\sin \theta_{e} |e^{+}\rangle
+\cos \theta_{e} |e^{-}\rangle
\]

Here we consider only the mixing separately of lepton (muon or
electron) charges, but not the flavour mixing as in oscillation
theories. In the maximum mixing ($\theta_{\mu, e} = \pi /4$)
Majorana leptons are fermions with a defined helicity  but
electrically neutral, then they may  travel a long way without
significant electro-magnetic interaction. At variance with
neutrinos, Majorana leptons are space-like particles with a rest
mass suggested almost equal to that of the corresponding Dirac
leptons. They may regenerate the Dirac component after a definite
period because of the lepton (or baryon) asymmetry in our
space-like frame. They may also be depolarised while traversing a
massive medium. During the process of regeneration or
depolarisation the lepton charge is changing, separately, for
Majorana electron and muon. However, we suggest that the total
lepton charge $L = L_{\mu} + L_e$ is conserved, that provides the
total electric charge unchanged. At the final moment when
particles are completely depolarised we may get the normal Dirac
components as:
\[
|\phi_{\mu}\rangle
=\frac{1}{\sqrt{2}}(|\phi_{\mu}(\mbox{Left})\rangle +
|\phi_{\mu}(\mbox{Right})\rangle)=|\mu^{-}\rangle;
\]
\[
|\phi_{e}\rangle =\frac{1}{\sqrt{2}}(|\phi_{e}(\mbox{Left})\rangle
- |\phi_{e}(\mbox{Right})\rangle)=|e^{+}\rangle.
\]

When the depolarisation time is less than the muon lifetime we may
observe not only a positron but also a muon as neutrino decay
products. However, if the regeneration period or depolarisation
lasts longer the muon lifetime, we may observe only a positron and
products of  the Dirac muon decay after regeneration or
depolarisation. Another possibility is that there is neither
regeneration nor depolarisation because of the conservation of
lepton charges and Majorana leptons remain almost sterile as dark
matter and we can never see them.

Along with the three-body decay there is a possibility of two-body
decay as an alternative mechanism. One can consider a
well-coupling singlet quasi-muonium consisted of muon-electron
pair under attractive Coulomb interaction at a short distance
($\geq 10^{-16}$ cm), which plays a similar role of the majoron in
the two-body decay as described in Ref.~\cite{barg}, however, it
does not need flavour mixing. The quasi-muonium, being
electrically neutral with a rest mass less than (or close to) the
total rest mass of the constituent particles,  has to decay into a
muon-electron pair after some period, let say, equal to the muon
lifetime. The existence of such a quasi-muonium, however, is
unlikely as it would be observed in different experiments  at
accelerators or in atmospheric cosmic rays. Obviously, the two-
and three-body decay mechanisms induce different energy spectra of
decay products, providing a means to identify the actual decay
mode.

In the next section we show that the three-body decay in the first
order approximation (without any free parameter) may give a
satisfactory interpretation of all short base-line oscillation
experiments at nuclear reactors as well at accelerators. As a next
step we predict some consequences of  this decay mode for long
base-line oscillation experiments.

\section{Interpretation of short base-line oscillation experiments}

We summarise here our results in Ref. \cite{bloi} as a
demonstration of the suggested mechanism to interpret short
base-line oscillation experiments carried out at accelerators
(including LSND, KARMEN and NOMAD), as well as at nuclear reactors
(including Bugey and the first short base-line Chooz).

\begin{figure}[h]
\begin{center}
\leavevmode
\includegraphics[width=0.7\columnwidth]{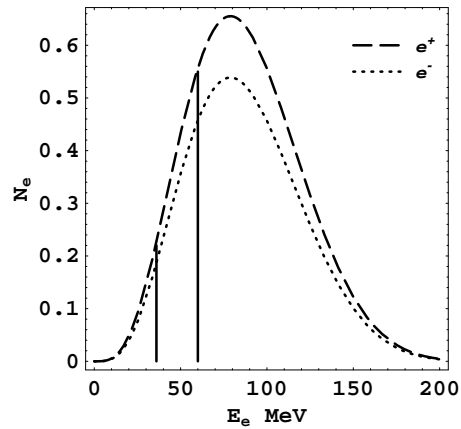}
\caption{Spectra of cc-events of LSND \label{fig:lsnd}}
\end{center}
\end{figure}

The positive effects of LSND have been interpreted as oscillations
of the muon anti-neutrino produced in muon decays at rest (DAR),
and oscillations of the muon neutrino produced in muon decays in
flight (DIF). In both cases muons are decay products of the pions
produced at the target A6. Instead of this oscillation
interpretation, in our Ref.~\cite{bloi} the LSND effects were
explained by the decay in flight of muon neutrinos mainly from
"minor" targets A1 and A2 but not by the decay at rest of those
from the major target A6. The last one (A6) contributed only about
20\% in the total effects of decay in flight. In
Fig.~\ref{fig:lsnd} are shown our calculated spectra of
cc-positrons and cc-electrons produced in interactions of electron
anti-neutrinos and neutrinos as products of muon neutrino decay
with hydrogen and carbon nuclei in the fiducial volume.

The calculated spectra are integrated over the range from $36$ to
$60$ MeV for cc-positrons and from $60$ to $200$ MeV for
cc-electrons, which give $9.9$ positrons and $37.5$ electrons.
Comparing with LSND data which give $22.0\pm 8.5\pm 8.0$
cc-positrons and $18.1\pm 6.6\pm 4.0$ cc-electrons, our results
are thus in an acceptable agreement with them.

In KARMEN as a sister project of LSND only decays at rest (DAR) of
pions and muons were taken into account, therefore all muon
neutrinos had an energy below threshold ($106.1$ MeV) and could
not decay.
\begin{figure}[h]
\begin{center}
\leavevmode
\includegraphics[width=0.7\columnwidth]{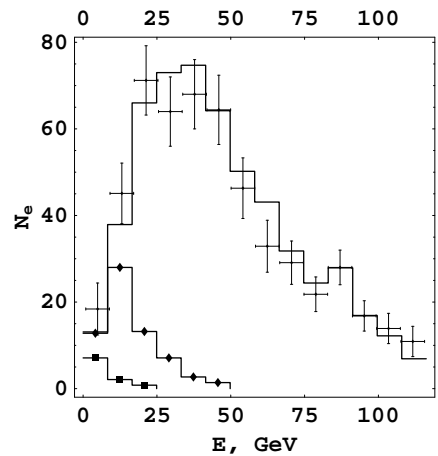}
\caption{Decay contribution in the cc-spectrum NOMAD
\label{fig:nomad}}
\end{center}
\end{figure}

We have calculated the neutrino decay at high energy in the NOMAD
experiment following the same procedure as for LSND. The fiducial
volume containing about $1.5\ 10^{30}$ nucleons and the initial
muon neutrino spectrum were taken from Ref.~\cite{alte}. The
$\nu$N-cross-section is proportional to energy as $0.78\ 10^{-38}$
$E_\nu$ (GeV) cm$^2$. In Fig~\ref{fig:nomad} we show the total
spectrum of cc-electrons, where the dash-line is a Monte-Carlo
simulation without oscillation; the calculated contribution of the
neutrino decay (box-line) consists only of $10.0$ events in the
range of $1-20$ GeV, namely $5$ times less than expected from
oscillation (star-line). This is to conclude that our decay
calculation is in a good agreement with the data of NOMAD.
However, the poor statistics and low energy resolution in the
lowest range of $1-20$ GeV do not allow  to separate the effect of
neutrino decay in flight (DIF) from the background.

Concerning electron (anti-) neutrinos, they are suggested stable
as time-like electrons (positrons), and certainly can not decay.
As a result, oscillation experiments at nuclear reactor producing
only electron anti-neutrinos could not see any  effect. The recent
attempt as a reactor long base-line project at Chooz is not an
exception \cite{apol}. For the solar anomaly concerning electron
neutrinos, we have proposed in Ref.~\cite{symp} a qualitative
interpretation by  a total depolarisation of the pseudo-spin of
time-like electrons passing through a thickness of dark plasma
matter. It could give in the first order approximation a roughly
50\% deficit of the expected solar neutrino flux which is in
agreement with the averaged experimental data $R = 0.46 \pm 0.06$.

\section{Estimation of muon neutrino decay at accelerator long base-line projects}

In the present section we deal with the three-body decay of muon
neutrinos at long base-line experiments, compared to oscillation
predictions. For this purpose we consider the projects: i/ K2K at
KEK-Super-K currently running \cite{jung}; ii/ MINOS at
Fermilab-Soudan,  already approved and in preparation \cite{mich}
and iii/ ICARUS at CERN-Gran Sasso waiting for approval
\cite{rubb}.

The decay in flight of muon neutrinos are calculated using the
same parameters as for muons (decay width $\Gamma$ or life-time
$\tau_\mu$, rest mass $m_0$) and the equation for decay
probability reads:
\begin{equation}
P_d = 1 - \exp\{-m_{0}(GeV)/(\tau_\mu c)\ L(km)/E_{\nu}(GeV)\};
\label{dprob}
\end{equation}

The two-neutrino oscillation probability is calculated using the
well-known formula:
\begin{equation}
P_{os} = \sin^{2}2\theta \ sin^{2}\{1.27\ \Delta m^{2}(eV^{2})\
L(km)/E_\nu (GeV)\}; \label{osprob}
\end{equation}

Here we use the most favourable parameters of the LSND \cite{atha}
for the ($\nu_\mu \rightarrow \nu_e$) oscillation
($sin^{2}2\theta$, $\Delta m^2$) $=$ ($6.10^{-3}$, $19 eV^2$) and
also those of the Kamiokande atmospheric anomaly \cite{fuku} for
the ($\nu_\mu \rightarrow \nu_\tau$) one: ($sin^{2}2\theta$,
$\Delta m^2$) $=$ ($0.95$, $5.9\times 10^{-3} eV^2$).
Table~\ref{tab:decay} reviews the estimation of decay and
oscillation probabilities of muon neutrinos from the long
base-line experiments at a mean energy averaged over all a
spectrum of each neutrino source. In this table we give not only
long base-line distances but also  the base-line of the companion
short distance detectors. Included are: i/ the KEK front detector;
ii/ COSMOS at Fermilab, and iii/ JURA at CERN.
\end{multicols}
\rule{8.4cm}{.1mm}\rule{-.1mm}{.1mm}\rule{.1mm}{2mm}
\begin{table}[h]
\caption{Calculation of decays and oscillations  probabilities of
muon neutrinos \label{tab:decay}} \vspace{0.4cm}
\begin{center}
\begin{tabular}{cccccc}
Project/Detector & $E_\nu$ (GeV) & $L$ (km) & $P_{os}$ LSND &
$P_{os}$ Atm. & $P_d$ \\ \hline K2K/Super-K & $1.5$ & $250$ & $1.\
10^{-3}$ & $0.89$ & $\approx 1.0$ \\ K2K/Front Det. & $1.5$ &
$300$ m & $3.4\ 10^{-5}$ & $2.1\ 10^{-6}$ & $0.02$\\
Fermilab/MINOS & $11.$ & $732$ & $8.6\ 10^{-3}$ & $0.22$ &
$\approx 1.0$ \\ Fermilab/COSMOS & $11.$ & $500$ m & $4.8\
10^{-3}$ & $1.1\ 10^{-7}$ & $7.2\ 10^{-3}$ \\ CERN/ICARUS & $30.$
& $730$ & $5.7\ 10^{-4}$ & $0.03$ & $0.98$ \\ CERN/JURA & $30.$ &
$17$ & $4.8\ 10^{-3}$ & $1.7\ 10^{-5}$ & $0.09$
\end{tabular}
\end{center}
\end{table}
\hspace{9cm}\rule{.1mm}{-4mm}\rule{.1mm}{.1mm}\rule{8cm}{.1mm}
\begin{multicols}{2}
We see from Table~\ref{tab:decay} that the LSND parameters produce
very small effects everywhere, while the Kamiokande parameters
give significant oscillation in the long base-line detectors, but
negligible effects in the companion short distance detectors.

\begin{figure}[h]
\begin{center}
\leavevmode
\includegraphics[width=0.7\columnwidth]{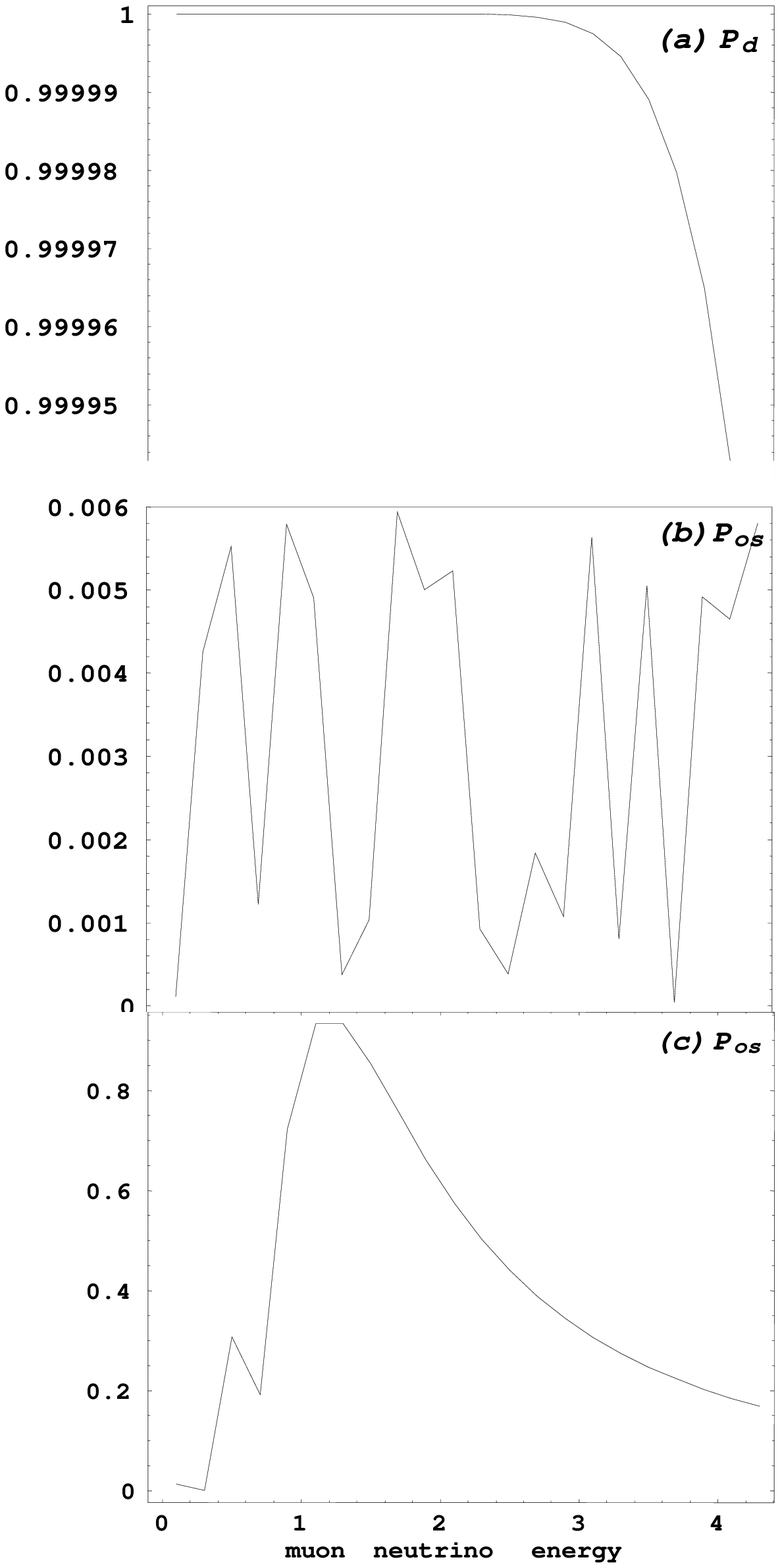}
\caption{K2K probabilities of decay (a) and oscillation: LSND- (b)
and Kamiokande- (c). \label{fig:k2kd}}
\end{center}
\end{figure}

The neutrino decay predicts very large effects on a far distance
that all long base-line detectors hardly observe muon-like events.
At Super-K and at MINOS almost all muon neutrinos have to decay
before reaching the detectors. Only about 2\% of the muon
neutrinos subsist at ICARUS. In Fig.~\ref{fig:k2kd} we illustrate
different behaviours of the decay (Fig.~3a) and oscillation
probabilities as functions of  muon neutrino energy in the K2K
long base-line project. For the oscillation we have taken again
parameters from both experiments LSND (Fig.~3b) and Kamiokande
(Fig.~3c). In the next presentation, we take into account of the
oscillation only the Kamiokande parameters.

\begin{figure}[h]
\begin{center}
\leavevmode
\includegraphics[width=0.7\columnwidth]{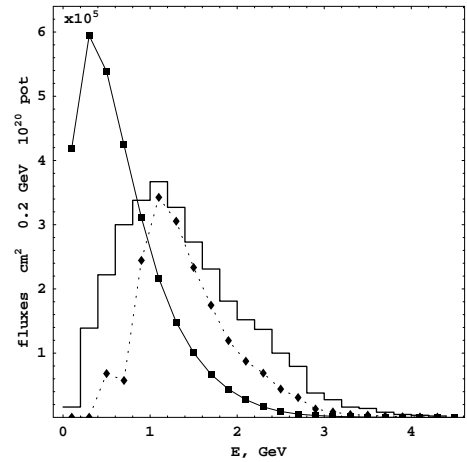}
\caption{K2K Muon and electron neutrino spectra.
\label{fig:muenu}}
\end{center}
\end{figure}

\begin{figure}[h]
\begin{center}
\leavevmode
\includegraphics[width=0.7\columnwidth]{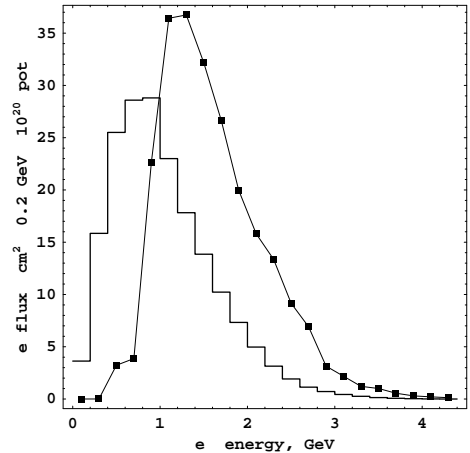}
\caption{K2K cc-electron spectra. \label{fig:cces}}
\end{center}
\end{figure}

Instead we may see muon-like events at short distance detectors,
where the amounts of electron neutrinos as neutrino decay products
may be significant, particularly at JURA, this proportion amounts
to 9\% of the total neutrino beam. At variance with oscillations,
electron neutrinos from neutrino decays have in average only  one
third of the initial muon neutrino energy as the calculated
electron neutrino spectrum (box-dash line) illustrated in Fig.~
\ref{fig:muenu} for K2K. For comparison there is shown a spectrum
(diamond-dot line) of tauon (or electron) neutrinos from the
$\nu_\mu \rightarrow \nu_\tau (\nu_e)$ oscillation, which repeats
the shape of the initial muon neutrino spectrum (histogram). In
Fig.~\ref{fig:cces} we show the corresponding calculated spectra
of electron-like events from the neutrino decays (histogram) and
from the oscillations (box-dash line). In the calculation we use
the $\nu_e$N cross-section as in \cite{kasu} for K2K, and a
simplified cross-section at higher energy for MINOS as $\sim C.\
10^{-38}$ $E_\nu$ (GeV) cm$^2$, where $C = 0.78$ for electron-like
events and $C = 0.62$ for muon-like ones. We see that the spectra
of electron-like events are very different each other for the
neutrino decay and for the oscillation. Particularly, the decay
spectrum is soften  significantly in variance with the oscillation
spectrum.




In Table \ref{tab:ratev} we summarise the total rates of cc-events
integrated over the calculated spectra of cc-events at K2K and
MINOS.
\end{multicols}
\rule{8.4cm}{.1mm}\rule{-.1mm}{.1mm}\rule{.1mm}{2mm}
\begin{table}[h]
\caption{Total rates of cc-events for $10^{20}$ protons on target
\label{tab:ratev}}
\begin{center}
\begin{tabular}{ccccc}
Project/Detector & Fiducial & osc. surviving & oscillation & decay
\\ & volume, kton & cc-muons & cc-electrons & cc-electrons \\
\hline K2K/Super-K & $22$ & $107.0$ & $235.6$ & $187.4$ \\
Fermilab/MINOS & $10$ & $7624.0$ & $1884.7$ & $5116.1$
\end{tabular}
\end{center}
\end{table}
\hspace{9cm}\rule{.1mm}{-4mm}\rule{.1mm}{.1mm}\rule{8cm}{.1mm}
\begin{multicols}{2}
For $10^{20}$ protons on target (p.o.t.) at K2K, if the muon
neutrino decay works, only $187$ cc-events will be collected and
if the particle identification (PID) works well only electron-like
events, but no muon-like events will be seen. Here we cannot
exclude the possibility that the electron neutrinos are Majorana.
In this case only the Dirac component may be detected, as in the
solar neutrino flux and, as a result, the total amount of
electron-like events may be decreased up to 50\% (for the maximum
mixing \cite{symp}).

For the oscillation mechanism there are different possibilities:
i/ when the ($\nu_\mu \rightarrow \nu_e$) version works, we may
have more electron-like events $(236)$ from the oscillation and an
amounts of muon-like ones $(107)$ from the survivors of the
initial muon neutrino spectrum; ii/ when the ($\nu_\mu \rightarrow
\nu_\tau$) version works, we may see the same amount of muon-like
events from the surviving muon neutrinos, however the rate of
cc-events from the tauon neutrinos decreases significantly, due to
the high threshold of the $\nu_\tau$N reaction and the low ratios
of tauon decay into muon or electron, which are 17.9\% and 17.4\%,
respectively \cite{mont}.

The first run of K2K at Super-K has  seen the first cc-event
\cite{suzu}. We have to wait for the next run in year 2000, for
sufficient statistics before making any definite conclusion.
According to Table~\ref{tab:ratev},  MINOS will give more
statistics to identify the origin of the oscillation hints.

\section{Conclusion}

In the Super-K long base-line detector  we expect to identify
muon-like events from electron-like ones as a criteria to test the
three-body decay  of  muon neutrinos. The shape of the spectra and
the quantity of cc-events are also sensitive to the origin of the
oscillation hints.

If the decay mechanism works, we suggest to the next long
base-line experiments to use intensively the shorter base-line
detectors. The optimal condition for observation is to get
significant amounts of neutrino decays to see electron-like
events, not to harm the intense flux to see also a proportion of
muon-like events, that a decay rate larger than say 20\% is
desirable. It leads to a ratio $L(km)/E_\nu (GeV) = 1.4$ and the
base-line $L$(km) equal to: 2.1; 15.3 and 41.8 km respectively,
for the KEK front detector, COSMOS and JURA. The experiments with
modified shorter base-line detectors might collect a significant
statistics for a short period. For example, 0.5 kton  fiducial
volume of the KEK front detector at the distance 2.1 km may see
$10^4$ muon-like and about (4 - 8)$\times 10^3$ electron-like
events for $10^{20}$ p.o.t.

\section*{Acknowledgement}

This work was funded by the Basic Research Program of the Ministry
of Science, Technology and Environment of Vietnam. We appreciate
Prof. P. Darriulat (CERN) for very useful discussion. We thank Dr.
Y. Yano (RIKEN), Prof. A. Masaike (Kyoto) and Dr. Y. Suzuki
(Super-K) for supporting the first author to attend the present
workshop.

\end{multicols}
\end{document}